\documentclass[aps,letterpaper,10pt,prl,reprint]{revtex4-1}
\usepackage{amsmath}
\usepackage{graphicx}

\usepackage{times}
\usepackage{mathptmx}

\newcommand{\figref}[1]{Fig.\;\ref{#1}}
\newcommand{\eqeqref}[1]{eq.\;\eqref{#1}}

\begin{document}
\title{Density matrix embedding: A simple alternative to dynamical mean-field theory}

\author{Gerald Knizia and Garnet Kin-Lic Chan}
\affiliation{Department of Chemistry, Frick Laboratory, 
Princeton University, NJ 08544, USA}


\begin{abstract}
We introduce DMET, a  quantum embedding theory for computing frequency-independent quantities, such as ground-state
properties, of infinite systems.
Like dynamical mean-field theory (DMFT), DMET maps 
the bulk interacting system to a simpler impurity model and is exact
in the non-interacting and atomic limits.
Unlike DMFT, DMET is formulated in terms of the frequency-independent local density matrix, rather than the local Green's function.
In addition, it features a finite, algebraically constructible bath of only one bath site per impurity site, 
 with no bath discretization error.
Frequency independence and the minimal bath make DMET a computationally simple and
 efficient method.
We test the theory in the 1D and 2D Hubbard models at and away from half-filling,
and we find that compared to benchmark data, total energies, correlation functions, and metal-insulator transitions are well reproduced, at a tiny computational cost.
\end{abstract}
\maketitle

Dynamical mean-field theory\cite{
Metzner89,Georges92,Georges96,Kotliar06,Maier05}
(DMFT) has developed into
a powerful embedding framework for bulk quantum systems.
Its central idea is to self-consistently map the infinite bulk system
onto an impurity model with only a few
 interacting {impurity} sites embedded in an infinite
non-interacting {bath}\cite{ Metzner89,Georges92,Georges96,Kotliar06,Maier05}.
In many settings, such impurity models can be solved using high-level many-body methods (so called impurity solvers),
\cite{GullRMP,Bulla98,Bulla08b,Garcia04,Nishimoto06,Caffarel94,hafermann:superperturbationimpuritysolver,Zgid11,zgid:cisolver}
owing to the small number of interactions.
Through the bath embedding, DMFT yields predictions that closely approach the bulk limit,
despite the greatly simplified treatment of interactions.

The basic quantum variable in DMFT is the local Green's function, $i g_{ij}(\omega)=\langle a^\dag_i [\omega-(H-E)]^{-1} a_j\rangle$.
As a function of frequency, it provides access to the local
 density of states as well as to static quantities such as energies. However, there are 
reasons to consider simpler 
frequency-independent quantum variables, too.  
For many applications  frequency information is not required; for example, energies
can be calculated from time-independent  states alone, as can
energy derivatives such as compressibilities or static correlation functions,
and many other properties.
Additionally, calculating stationary states, such as the ground-state, is much easier than
calculating Green's functions, not the least due to the practical availability of many powerful
numerical techniques (e.g., the density matrix renormalization group and its tensor
network extensions \cite{white92,Verstraete08}, coupled cluster and configuration
interaction theories \cite{Helgaker00}, and 
lattice diffusion \cite{trivedi1989green}, auxiliary field \cite{zhang2003quantum}, and variational
Monte Carlo \cite{sorella2005wave,neuscamman:cpsoptimization}).
An embedding framework based on a frequency-independent  variable 
offers a potentially more efficient as well as a more flexible route to access  static properties of bulk systems,
including the possibility of using ground state methods as impurity solvers.

Here, we propose a {\it density matrix embedding theory} (DMET) with
the following features: (i) the infinite bulk problem is mapped onto a self-consistent impurity problem, consisting of interacting impurity and non-interacting bath sites,
(ii) the single-particle density matrix $\langle a^\dag_i a_j\rangle$
is the quantum variable, rather than the Green's function, and no frequency-dependent
quantities appear in the theory, (iii) the bath representation consists of a single bath site per impurity site
(which is sufficient to exactly capture embedding effects at the mean-field level) and (iv)
 the bath can be constructed algebraically without any fitting. Feature (i) is analogous to DMFT, and as we show
below, the basic physics of DMET is similar to DMFT. Features (ii)-(iv), however, are different. They allow for the primary numerical advantage of DMET: computing ground-state properties of a cluster
model with  $L$ impurity sites
requires  only solving for   the ground-state of a cluster plus bath problem of size $2L$, and this is much cheaper
than the corresponding DMFT calculation. 

To motivate the DMET construction, we first consider an exact single-site
embedding of the infinite lattice Hubbard model, with Hamiltonian
\begin{align}
H=\sum_{\langle ij\rangle \sigma} t a^\dag_{i\sigma} a_{j\sigma} + \sum_i U n_{i\uparrow} n_{i\downarrow}.\label{eq:FullLatticeH}
\end{align}
For simplicity, we focus on ground-state properties, but we outline the extension to excited and thermal states below.
The ground-state $|\Psi\rangle$ of $H$ can be mapped onto the ground state of a simple impurity model, consisting of a single impurity site embedded with a single bath site and
with Hamiltonian $H^\prime$.
This follows from the Schmidt decomposition of $|\Psi\rangle$, $|\Psi\rangle=\sum_i^M \lambda_i |\alpha_{i}\rangle |\beta_{i}\rangle$
\cite{peschel:EntangleMentInSolvableManyParticleModels},
where $|\alpha_{i}\rangle$ are states of a single Hubbard site (viewed as an ``impurity'') and $|\beta_{i}\rangle$ are states
in the Hilbert space of the remaining lattice sites. Note that the number of $|\beta_{i}\rangle$ states, $M$,
equals the number of states of the single impurity site. Consequently, $|\beta_{i}\rangle$  {\it can be interpreted as the
states of a single bath site}. With this identification, the exact impurity Hamiltonian  $H^\prime$ (with the same ground-state as $H$) may be constructed
 by projecting $H$ onto the Schmidt basis of impurity and bath states, 
$H^\prime = \sum_{iji^\prime j^\prime} |\alpha_i \beta_j\rangle \langle \alpha_i \beta_j|H |\alpha_{i^\prime} \beta_{j^\prime}\rangle \langle \alpha_{i^\prime} \beta_{j^\prime}|$. More explicitly, $H^\prime$ can be expressed using the 
impurity and bath fermionic operators $a^{(\dag)}$, $b^{(\dag)}$, 
\begin{align}
H^\prime &= c+\sum_\sigma  \epsilon_{\beta} b^\dag_\sigma b_\sigma 
+ v (a^\dag_\sigma b_\sigma + b^\dag_\sigma a_\sigma) \notag\\
&+ \sum_{\sigma \neq \sigma^\prime}   v^\prime m_{\sigma}
   (a^\dag_{\sigma^\prime} b_{\sigma^\prime} + b^\dag_{\sigma^\prime} a_{\sigma^\prime}) 
 + U n_{ \uparrow} n_{\downarrow} + U_\beta m_{\uparrow} m_{\downarrow}
\label{eq:two_site_map}
\end{align}
where $n_\sigma=a^\dag_\sigma a_\sigma$, $m_{\sigma} = b^\dag_\sigma b_\sigma$, and $c,\epsilon_\beta, v, v', U_\beta$ follow from matrix elements of $H$ with the Schmidt basis,
for example, $v=\langle \alpha_0 \beta_0 | H | \alpha_1 \beta_1\rangle$.
Because the impurity  Hamiltonian $H^\prime$ has the same ground-state $|\Psi\rangle$ as the Hubbard Hamiltonian $H$,
 expectation values of the infinite lattice
can be obtained {\it exactly} from the impurity model.  (This is very different from DMFT,
where expectation values of the lattice and impurity model are not related in a simple way).
For example, the local density matrix is given by  $\langle a^\dag_\sigma a_\sigma \rangle_{H^\prime} \equiv \langle a^\dag_{0\sigma} 
a_{0\sigma}\rangle_H$ (the latter referring to lattice sites, \eqeqref{eq:FullLatticeH}), while
the lattice Hubbard energy  per site, $E$, is obtained from the terms in $H^\prime$ which contain the impurity $a^{(\dag)}$ operator,
$ 
E \equiv  \langle \sum_\sigma v  (a^\dag_{\sigma} b_\sigma + b^\dag_{\sigma} a_\sigma)
 + \sum_{\sigma \neq \sigma^\prime}v^\prime  m_{\sigma}
   (a^\dag_{\sigma^\prime} b_{\sigma^\prime} + b^\dag_{\sigma^\prime} a_{\sigma^\prime})
+ U n_{ \uparrow} n_{ \downarrow}\rangle_{H^\prime}
$.

The above shows that an impurity model with a single impurity site and a single bath site is in principle sufficient to exactly represent ground state properites.
However, the exact construction is
not practically useful: the bath terms in $H^\prime$ require knowledge of the interacting ground-state $|\Psi\rangle$ on the infinite lattice and its Schmidt decomposition.
The basic idea in DMET is to replace the exact embedding  of the Hamiltonian $H$  by one that is {\it exact for a one-particle  mean-field lattice Hamiltonian} $h$.
The corresponding mean-field embedding bath terms are then easy to compute because the ground-state of $h$ is a Slater determinant $|\Phi\rangle$, and
its Schmidt decomposition is easily obtained, {\it at a cost no greater than the one-particle diagonalization of $h$ itself} \cite{klich2006lower,peschel:EntangleMentInSolvableManyParticleModels,dmet:hp}.
As $h$ we choose the one-particle part of $H$ combined with an on-site mean-field interaction operator $u$ (to be determined),
\begin{align}
h=\sum_{\langle ij\rangle \sigma} t a^\dag_{i\sigma}a_{j\sigma}+ \sum_{i\sigma} u (n_{i\uparrow} + n_{i\downarrow}).  \label{eq:h}
\end{align}
The corresponding DMET impurity Hamiltonian $H_\text{imp}$ contains a single interacting impurity site, now embedded with a {\it non-interacting} bath site,
\begin{align}
H_{\text{imp}} = \sum_\sigma \left(v (a^\dag_{\sigma} b_\sigma + b^\dag_\sigma a_{\sigma}) + {\tilde u}\,b^\dag_\sigma b_\sigma\right) + U n_{\uparrow} n_{\downarrow}. \label{eq:Himp}
\end{align}
The terms involving the bath operators $b^{(\dag)}$ in $H_{\text{imp}}$ are constructed analogously to the bath terms in \eqeqref{eq:two_site_map},
by projecting $h$ onto the Schmidt basis of $|\Phi\rangle$, e.g. given $|\Phi\rangle = \sum_{i}^m \lambda_i |\alpha_i\rangle |\beta_i\rangle$,
$v= \langle \alpha_0 \beta_1 | h | \alpha_1 \beta_0\rangle$. Note that the bath terms arising in this way contain only
one-particle operators. 
The mean-field interaction operator $u$, which defines both the lattice Hamiltonian $h$ and (indirectly) the impurity Hamiltonian $H_\text{imp}$,  is analogous to the 
DMFT self-energy. By changing $u$, we also change the bath terms in $H_\text{imp}$, and  we can adjust $u$ to obtain a mean-field
embedding that ``optimally'' mimics the exact embedding. One way is to require full self-consistency of the local density matrix, that is, that
 the mean-field lattice Hamiltonian $h$ and the DMET impurity Hamiltonian $H_\text{imp}$ yield the same local density matrix:
$\langle a^\dag_{\sigma} a_{\sigma} \rangle_{H_{\text{imp}}} = \langle a^\dag_{0\sigma} a_{0\sigma} \rangle_{h}.$ This is similar to
 the self-consistency condition on the Green's function in DMFT.
We here use a slightly different self-consistency condition, which we have found to be numerically favorable. We minimize
the difference between the total density matrices (using both impurity and bath operators) evaluated
for the ground states of $h$ and $H_{\text{imp}}$,
\begin{align}
\min_{u} \sum_{\sigma, c \in \{a, b\} } \left(\langle c^\dag_\sigma c_\sigma \rangle_{H_\text{imp}} -
\langle c^\dag_\sigma c_\sigma \rangle_{h}\right)^2.
\end{align}
This approximately maximizes the overlap between the mean field wave function and the full wave function in the impurity model.
Other choices are possible, similar to the different choices of self-consistency condition in DMFT and self-energy functional theories \cite{Potthoff03,Maier05}.

In the exact embedding construction, the impurity Hamiltonian $H^\prime$ and
the infinite Hubbard Hamiltonian $H$ shared the same ground-state, thus exact expectation
values could be obtained from the impurity model.
In the case of the DMET impurity Hamiltonian $H_\text{imp}$, this is no longer the case, because
the embedding is constructed for the mean-field $h$ rather than $H$. Nonetheless,
in DMET we make the approximation that lattice quantities are  approximated by
the expectation values of the impurity model. The local density matrix in DMET is thus
 defined as $\langle  a^\dag_{\sigma} a_{\sigma} \rangle_{H_{\text{imp}}}$,
and the energy-per-site is
\begin{align}
E= \sum_\sigma v\langle a^\dag_{\sigma} b_\sigma + b^\dag_{\sigma} a_\sigma\rangle_{H_{\text{imp}}} + U\langle n_{\uparrow} n_{\downarrow}\rangle_{H_{\text{imp}}}.
\end{align}

So far we have discussed single-site DMET. Cluster
extensions can also be formulated.
The simplest cluster extension of DMET for $L$ impurity sites is 
analogous to $L$ site cellular DMFT \cite{Maier05}, and the
impurity Hamiltonian $H_\text{imp}$ is given by
\begin{align}
H_{\text{imp}} = \sum_\sigma \sum_{ i,j=1}^{L} \left(v_{ij} (a^\dag_{i\sigma} b_{j\sigma} + b^\dag_{i\sigma} a_{j\sigma}) +{\tilde u}_{ij} b^\dag_{i\sigma} b_{j\sigma}\right) + U n_{i\uparrow} n_{i\downarrow} 
\end{align}
where  the  bath $b_i^{(\dag)}$, interaction $u_{ij}$, and coupling $v_{ij}$ operators are all generalized to $L$ sites. The construction
of the couplings from the Schmidt basis of $h$, the self-consistency conditions on $u$, and the approximation of lattice quantities
by impurity quantities, are all directly analogous to the single-site case.
Note that this kind of cluster DMET breaks translational invariance, similarly to cellular DMFT \cite{Maier05}. 
\begin{figure*}
   \centering
   (a){\includegraphics[width=3in]{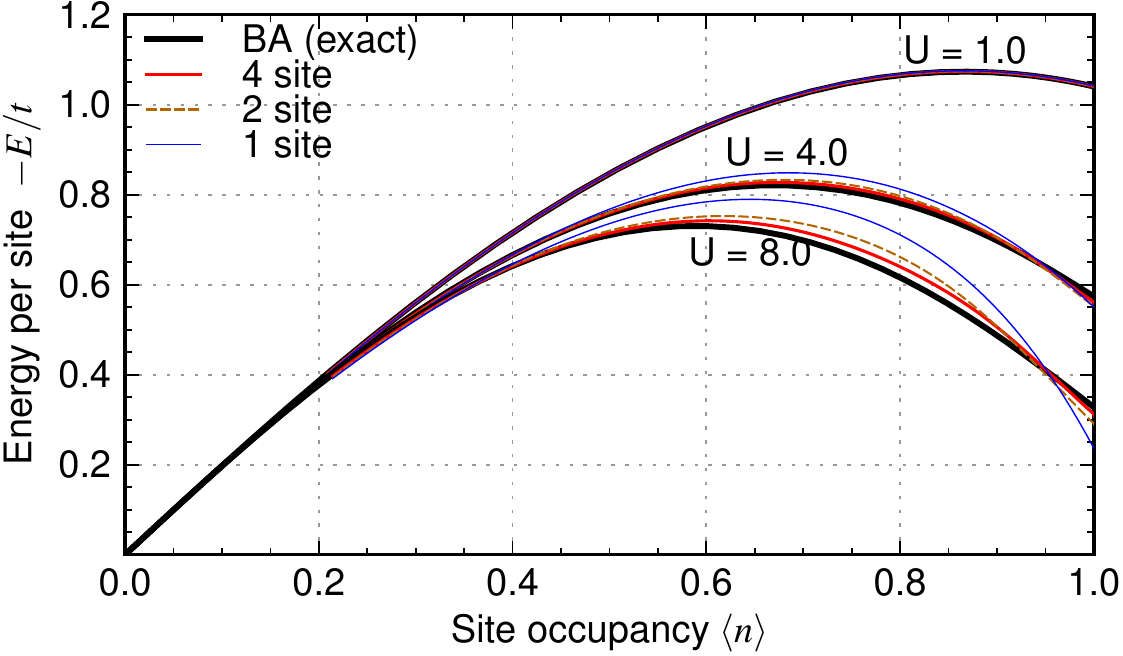}}
   \hspace*{1em}
   (b){\includegraphics[width=3in]{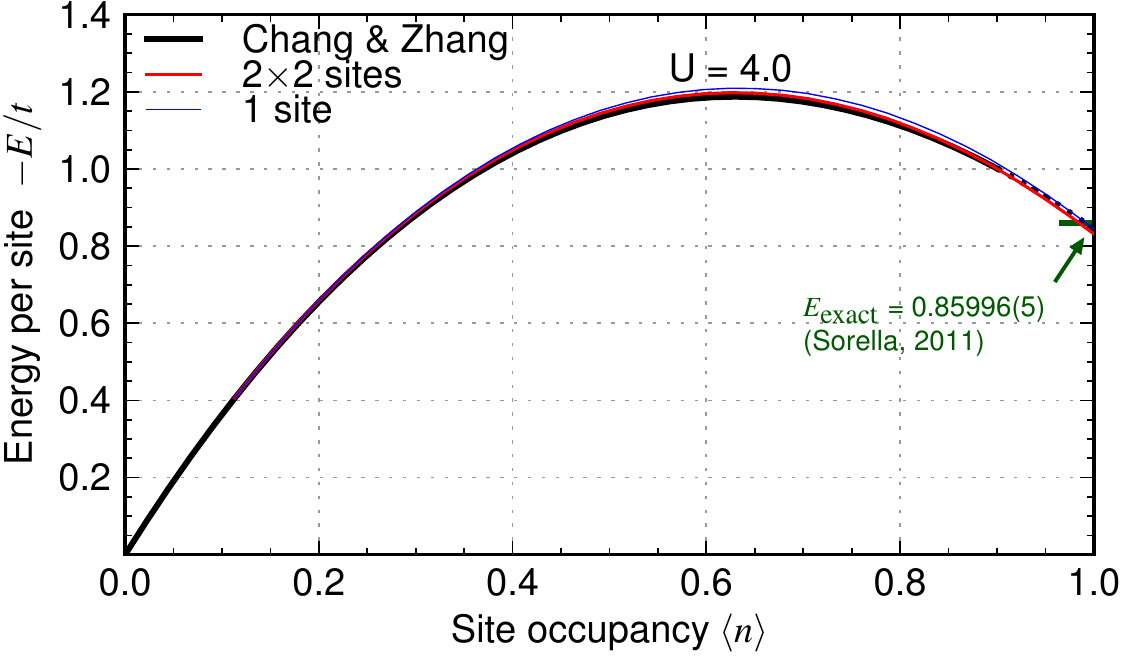}}
   \caption{(color online) $E(n)$ with various
      cluster sizes: (a) 1D Hubbard model, and (b) 2D Hubbard model.
      In both cases good agreement with reference data is found. For (b), QMC reference data is taken from
      Refs. \onlinecite{zhang:hub2dref_afmc,sorella:hub2ref_u4hlf}.}
   \label{fig:E_vs_n}
\end{figure*}

Further extensions are possible, but not followed here. While we have focused on
ground-states, the DMET construction is equally
applicable to excited states (by carrying out the Schmidt decomposition for
the excited state, rather than the ground-state). In the case of thermal states, we would regard $\exp(-\beta H)$
as a state vector in the enlarged Liouville space to carry out the Schmidt decomposition, leading to impurity
and bath sites with twice the number of degrees as in the case of pure states.

How does DMET perform? Similarly to DMFT, the DMET construction is
exact in the non-interacting ($U=0$) and atomic ($t=0$) limits of the Hubbard model \cite{Georges96}.
In the non-interacting limit, the mean-field lattice Hamiltonian and Hubbard Hamiltonian are the same, $h=H$, thus
the mean-field embedding used in DMET is exact. In the atomic limit, the Hubbard Hamiltonian decouples the sites, the impurity-bath coupling $v=0$ vanishes,
and  the impurity expectation values are  identical to  Hubbard model expectation values.
DMET thus acts similarly to DMFT in providing  an interpolation between metallic and Mott insulating behaviour.
Note also that DMET has the same diagrammatic structure as DMFT; solution of the impurity Hamiltonian with an exact local
interaction ensures that the resulting observables have diagrams where the local interactions are treated to all orders. The
different self-consistency condition and bath construction, however, mean that the theories are not identical even in the static limit.

To assess the validity and accuracy of DMET away from exact limits,
we now turn to numerical DMET studies of the 1D and 2D Hubbard models, as a function of $U$ and filling.
These models were chosen because high-quality reference data is available.
In 1D, we compare to exact results from the Bethe
ansatz\cite{lieb:BetheAnsatz,shiba:BetheNonhalfFilling}, while in
2D we compare to recent auxiliary field quantum Monte
Carlo (QMC) calculations\cite{zhang:hub2dref_afmc,sorella:hub2ref_u4hlf}.
In the numerical DMET calculations, the infinite lattice used to define the mean-field Hamiltonian $h$ in \eqeqref{eq:h}
is approximated by a large lattice with antiperiodic boundary conditions, using 480 sites in 1D case, and 24$\times$48 sites in the
2D case. The resulting finite size errors are substantially smaller than the intrinsic errors from the DMET approximation. 
 For the DMET impurity solver 
we employed an exact diagonalization algorithm\cite{knowles:fci}. The full source code for the prototype DMET and the impurity solver is freely available\cite{dmet:hp}.

\begin{figure}[t]
   \centering
   \includegraphics[width=3in]{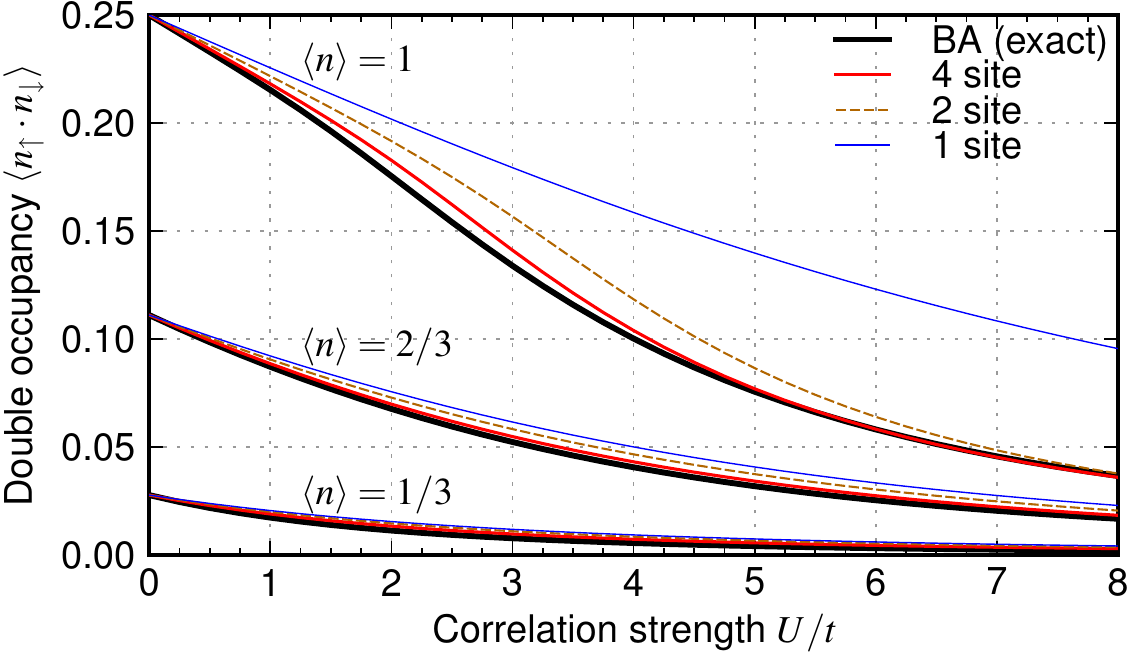}
   \caption{(color online) 1D Hubbard model: Local double occupancy $\langle n_\uparrow^i\cdot n_\downarrow^i\rangle$ as
      function of $U$; at half-, $2/6$, and $1/6$ filling. Except for the 1-site DMET case, the reference shapes are well reproduced.}
   \label{fig:U_vs_SzSz}
\end{figure}

We first discuss DMET's ability to describe energetics and correlation functions in the 1D and 2D
Hubbard models. 
\figref{fig:E_vs_n} gives the energy per site $E$, as a function of site occupancy $n$, using
 DMET of various cluster sizes.
Note that while the 1D Bethe ansatz results are exact, the 2D AFQMC reference data is not \cite{zhang:hub2dref_afmc}, except at half-filling \cite{sorella:hub2ref_u4hlf}.
As seen from the figure, the overall shapes of the reference $E(n)$ curves are reproduced well. Unsurprisingly, DMET is most accurate 
 at smaller $U$. The accuracy increases with impurity cluster size, but even 
single-site DMET produces reasonable energies: at half-filling in 1D, single-site DMET is accurate to 0.12\% at $U=1$, and $4\%$ at $U=4$,
while at half-filling in 2D, it is accurate to $1.8\%$ at $U=4$. 
In \figref{fig:U_vs_SzSz} we plot the local double-occupancy $\langle n_\uparrow n_\downarrow\rangle$, which is a measure of the Mott insulating character of the state \cite{Parcollet04}. DMET also gives a good description of this quantity, although at half-filling,
single-site DMET does not  capture the correct change in curvature of $\langle n_\uparrow n_\downarrow\rangle$ as a function of $U$  due to the 
neglect of short-range singlet formation, while larger clusters recover this interaction effect.

\begin{figure}
   \centering
   \includegraphics[width=3in]{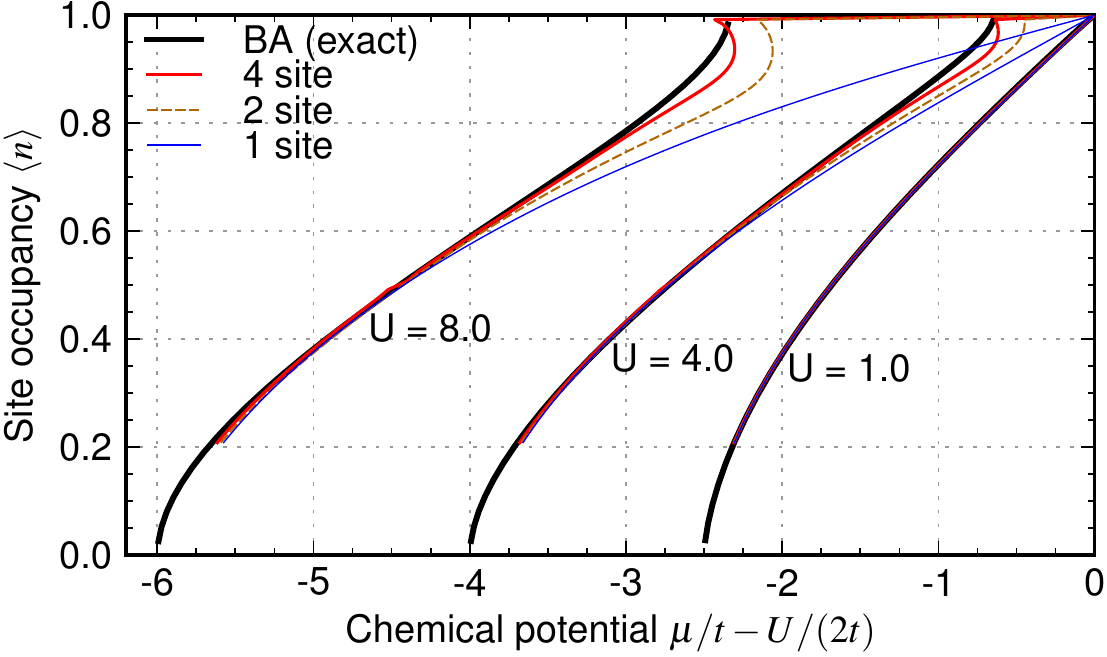}
   \caption{(color online) Metal-insulator transition in the 1D-Hubbard model: a Mott
   gap is seen for cluster sizes $> 1$. The size of the gap is accurately reproduced.}
   \label{fig:MIhub1}
\end{figure}

\begin{figure}
   \centering
   \includegraphics[width=2.5in]{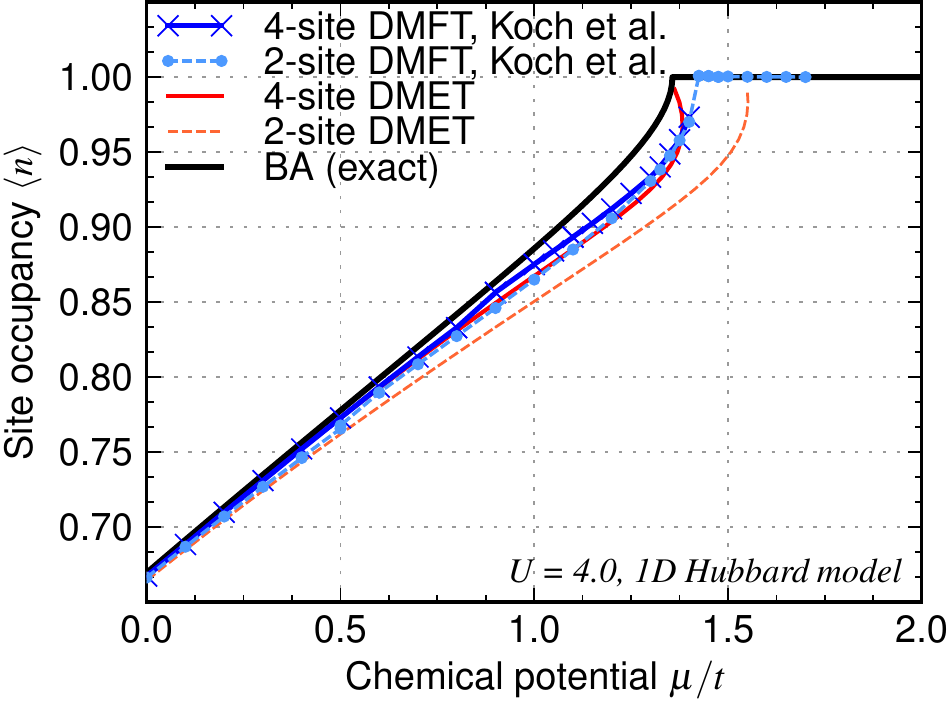}
   \caption{(color online) Comparison of DMET and DMFT descriptions of the
      Metal-insulator transition. DMFT data is taken
      from Ref.~\onlinecite{Koch:sumrules}.}
   \label{fig:MIhub1Dmft}
\end{figure}

\begin{figure}[t]
   \centering
   \includegraphics[height=1.36in]{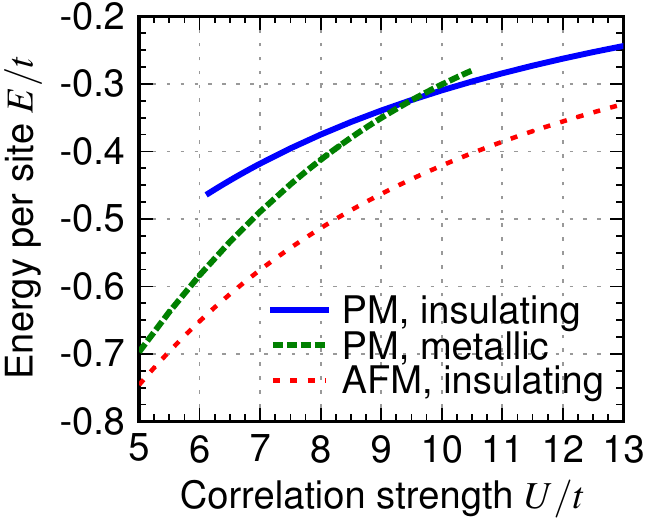}
   \hspace*{0.01in}
   \includegraphics[height=1.36in]{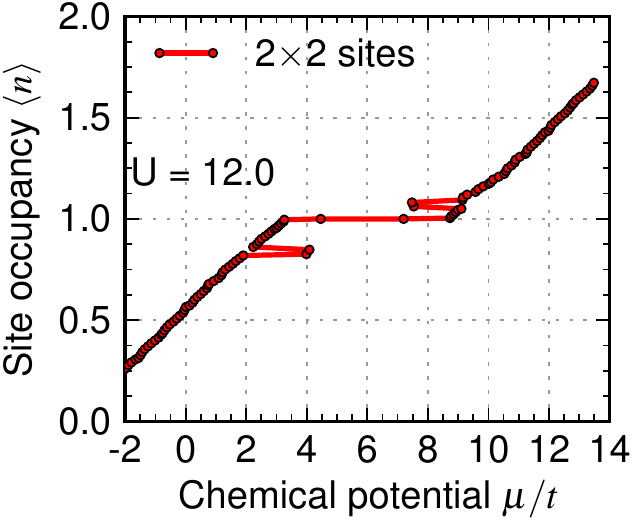}

   \caption{(color online) 2D Hubbard model: (left) Phase stability of metallic
      and insulating solutions at half filling as function of $U$;
      (right) Paramagnetic metal-insulator transition at $U=12$.}
   \label{fig:hub2d_metallic_insulating_phases}
\end{figure}

A central question in the DMFT treatment of the 1D and 2D Hubbard models is the
location and nature of the metal-insulator transition and the corresponding size of the Mott gap \cite{Park:CDMFTOfTheMottTransition,Sordi:UMuTPhaseDiagramHub2d,Balzer:MottHub2dVariationalPlaquette}.
In the antiferromagnetic DMFT solution a gap opens at $U=0$ due to perfect nesting. The paramagnetic DMFT solution is more complicated,
and there is
%Mott transition in the paramagnetic mean-field solution, and the corresponding size
 a critical interaction strength $U$, which depends on the cluster size, beyond which a Mott insulator
paramagnetic solution is stabilised. In a ground-state calculation, the opening of a Mott gap
can be detected by a vanishing compressibility, $d\langle n\rangle/d\mu$. 
In \figref{fig:MIhub1} we plot $\langle n\rangle$ against $\mu$ for DMET calculations
on the 1D Hubbard model. In the antiferromagnetic DMET calculations, a gap opens at $U=0$ as in DMFT.
In the paramagnetic DMET calculations,  a Mott transition is observed
in the cluster DMET calculations for larger $U$'s, although not in the single-site DMET. Note that single-site DMFT also
does not display a Mott transition at moderate $U$ \cite{capone:ClusterDmftMottHub1d}. Compared to the Bethe ansatz, the Mott gap
is well reproduced, with the error decreasing with cluster size. 
The $n(\mu)$
behavior close to half-filling shows a bi-stability, indicating that the 
metal-insulator transition is likely to be
 first-order, similar to what is found in DMFT. 
As seen from \figref{fig:MIhub1Dmft}, the shape of the $n(\mu)$ curves resembles that obtained in
recent cellular DMFT studies\cite{Koch:sumrules,capone:ClusterDmftMottHub1d}.
In the 2D Hubbard model, using a $2\times 2$ cluster DMET, we find
co-existence of metallic and insulating paramagnetic solutions over a range of $U$,
in addition to the antiferromagnetic solution (which is correctly identified as the lowest energy phase). The co-existence
region starts at about $U\approx 6$ and extends to $U\approx
10.5$. However, the metallic solution is
clearly favored up to $U\approx 9.5$ as seen from
\figref{fig:hub2d_metallic_insulating_phases} (left). Cluster DMFT \cite{Park:CDMFTOfTheMottTransition,Sordi:UMuTPhaseDiagramHub2d}
and variational plaquette \cite{Balzer:MottHub2dVariationalPlaquette} calculations also show
a significant co-existence region for paramagnetic metallic and insulating solutions
with the metal-insulator transition occurring near  $U \approx 6$. 
In \figref{fig:hub2d_metallic_insulating_phases} (right) we plot $n(\mu)$ for $U=12$. We see
that the structure away from half-filling is more complicated than in the 1D plots.
Note that this behavior occurs at moderate doping where complex phases of the Hubbard
model are expected to exist.
Overall, we find that
the DMET and DMFT descriptions of the paramagnetic metal-insulator transition are similar, but the DMET calculations
of ground-state properties
are much cheaper: each one requires computing only
the ground-state on a very small number of sites, which takes only seconds.

To summarize, here we introduced a density matrix embedding theory (DMET) to
obtain properties of infinite bulk interacting systems from a simpler
 embedded quantum impurity model. While similar to DMFT, DMET
is formulated in terms of the frequency independent local density matrix. The absence of
frequency yields both formal and practical advantages. Further, the DMET bath construction
requires only a small number of sites (one per impurity), and is obtained algebraically without non-linear fitting. There is no bath discretization error.
The simple bath construction and lack of frequency means that DMET calculations for static properties are \emph{much} simpler and faster than the
corresponding DMFT calculations. We showed that DMET contains similar local physics and is exact in the same limits as DMFT, namely
for weak interactions and weak couplings. Furthermore, for the 1D and 2D Hubbard models,
accurate behaviour for the energetics, correlations, and  metal-insulator transition was observed. 
Consequently, we conclude that DMET provides an appealing  alternative to DMFT for the
ground-state properties of infinite interacting systems. Applications to further problems, as well as extensions e.g.
to long-range interactions, are now underway.

Support was provided from the US Department of Energy, Office of Science, through the Computational Materials Science Network. We acknowledge
helpful discussions with A. Millis, D. Reichman, and C. Marienetti.

% \bibliography{refs,a1,fci_impuritysolve,embedding}
%merlin.mbs apsrev4-1.bst 2010-07-25 4.21a (PWD, AO, DPC) hacked
%Control: key (0)
%Control: author (8) initials jnrlst
%Control: editor formatted (1) identically to author
%Control: production of article title (-1) disabled
%Control: page (0) single
%Control: year (1) truncated
%Control: production of eprint (0) enabled
%

\end{document}